\documentclass[conference]{IEEEtran}
\IEEEoverridecommandlockouts

\usepackage{cite}
\usepackage{amsmath,amssymb,amsfonts}
\usepackage{algorithmic}
\usepackage{graphicx}
\usepackage{textcomp}
\usepackage{xcolor}
\usepackage{multirow}

\def\BibTeX{{\rm B\kern-.05em{\sc i\kern-.025em b}\kern-.08em
    T\kern-.1667em\lower.7ex\hbox{E}\kern-.125emX}}
\begin{document}

\title{CSI-MAE: A Masked Autoencoder-based \\ Channel Foundation Model\\

\author{Jun Jiang\IEEEauthorrefmark{1}\IEEEauthorrefmark{3}, Xiaolong Ruan\IEEEauthorrefmark{2}\IEEEauthorrefmark{3}, 
and Shugong Xu\IEEEauthorrefmark{2}\IEEEauthorrefmark{4}\\
\IEEEauthorrefmark{1}School of Communication and Information Engineering, Shanghai University, Shanghai, China \\
\IEEEauthorrefmark{2}Xi'an Jiaotong-Liverpool University, Jiangsu, China \\
Email: jun\_jiang@shu.edu.cn, Xiaolong.Ruan25@student.sjtlu.edu.cn, 
shugong.xu@xjtlu.edu.cn \\
\IEEEauthorrefmark{3} Co-first author
\IEEEauthorrefmark{4} Corresponding author
}
}

\maketitle

\begin{abstract}

Self-Supervised Learning (SSL) has emerged as a key technique in machine learning, tackling challenges such as limited labeled data, high annotation costs, and variable wireless channel conditions. It is essential for developing Channel Foundation Models (CFMs), which extract latent features from channel state information (CSI) and adapt to different wireless settings. Yet, existing CFMs have notable drawbacks: heavy reliance on scenario-specific data hinders generalization, they focus on single/dual tasks, and lack zero-shot learning ability. In this paper, we propose CSI-MAE, a generalized CFM leveraging masked autoencoder for cross-scenario generalization. Trained on 3GPP channel model datasets, it integrates sensing and communication via CSI perception and generation, proven effective across diverse tasks. A lightweight decoder finetuning strategy cuts training costs while maintaining competitive performance. Under this approach, CSI-MAE matches or surpasses supervised models. With full-parameter finetuning, it achieves the state-of-the-art performance. Its exceptional zero-shot transferability also rivals supervised techniques in cross-scenario applications, driving wireless communication innovation.

\end{abstract}


\section{Introduction}

Self-Supervised Learning (SSL) has emerged as a key paradigm in machine learning, firmly establishing itself as a cornerstone within the Artificial Intelligence (AI) domain \cite{gui2024survey}. In the rapidly evolving landscape of wireless communication, traditional supervised learning methodologies are confronted with formidable challenges. These challenges primarily stem from the dearth of labeled data, exorbitant annotation expenses, and limited scene generalization capabilities, all of which impede the development of robust and adaptable models for wireless communication systems. SSL presents a promising solution by enabling models to extract knowledge from unlabeled data, effectively capitalizing on the rich information embedded within channel state information (CSI). 

\begin{figure}[tbp]
\centering
\includegraphics[width=0.75\linewidth]{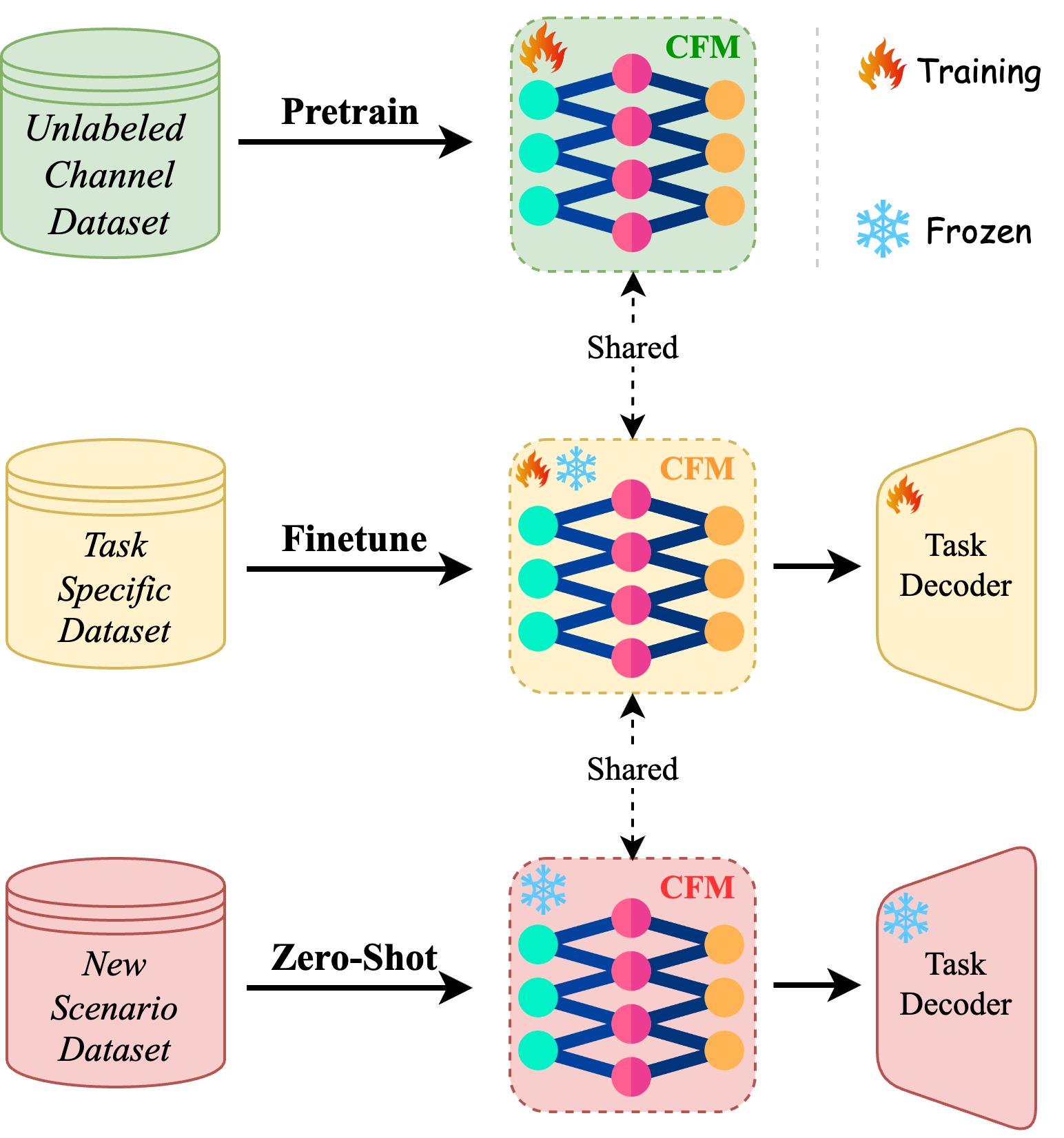}
\caption{The overall framework for CFMs.}
\label{fig:overall}
\end{figure}

The effectiveness of these SSL paradigms in learning general representations across diverse domains has spurred researchers to explore their applicability in specialized fields. Specifically, the unique challenges and data characteristics of wireless communication, such as  the diverse scenarios and tasks, present an ideal scenario for leveraging SSL. This has led to a growing interest in Channel Foundation Models (CFMs)\cite{csi-clip,cfmsurvey}, which aim to discover universal channel representations. CFMs learn transferable and generalizable CSI representations through pretraining, as shown in Fig.~\ref{fig:overall}. This approach significantly improves cross-task and cross-scenario generalization as well as data efficiency.

SSL aims to extract generalizable and transferable representations from large-scale unlabeled data, and its success can be attributed to two primary paradigms: Masked X Modeling (MXM) and Contrastive Learning (CL). MXM techniques rely on reconstructive inference, training models to predict missing components from incomplete data \cite{bert,mae,ssnet}. This design effectively captures semantic and structural priors, facilitating the generation of context-aware representations. For instance, in the field of Natural Language Processing, BERT \cite{bert} employs bidirectional self-attention mechanisms to predict masked tokens, while in computer vision, MAE \cite{mae} utilizes an asymmetric encoder-decoder architecture to reconstruct masked image patches during the pretraining phase. Conversely, CL, such as MoCo \cite{moco} and SimCLR \cite{simclr}, focus on discriminative feature learning. These methods maximize the similarity between augmented views of the same data instance while minimizing the similarity to other instances, thereby producing robust, domain-invariant representations that enhance generalization without the need for explicit supervision. Notably, recent studies \cite{dinov2,ibot} integrate MXM with CL to exploit their complementary strengths. These hybrid methods combine MXM's context reconstruction with CL's discriminative power, enabling fine-grained semantic capture through reconstruction while enhancing instance discrimination. The resultant representations are context-rich and highly discriminative, improving performance on various downstream tasks and enhancing self-supervised model generalization.

Several pioneering studies \cite{lwm,bert4mimo,wimae,wifo,wirelessgpt} using Masked Channel Modeling (MCM) have laid the foundation for the development of CFMs. LWM \cite{lwm} introduced the first channel-aware foundation model using, focusing on channel identification and Sub-6G-to-mmWave beam prediction. BERT4MIMO \cite{bert4mimo} adapted the BERT architecture to large-scale MIMO systems for high-dimensional CSI reconstruction. WiMAE \cite{wimae} incorporated the asymmetric encoder-decoder design of MAE to facilitate efficient and robust channel representation learning. WiFo \cite{wifo} proposed a unified spatio-temporal-frequency CFM for cross-configuration channel prediction, eliminating the need for task-specific finetuning.

Notwithstanding these achievements, existing CFMs are confronted with three significant limitations that are consistent with the broader challenges in wireless communication:

\begin{enumerate}

\item  Dependence on Scene-Specific Datasets. The majority of models utilize ray tracing data (e.g., DeepMIMO \cite{deepmimo}, SionnaRT \cite{sionna}) collected from specific environments. This results in limited generalization capabilities when compared to the standardized statistical channel models specified in 3GPP TR 38.901 \cite{38_901}, such as the Rural Macrocell (RMa), Urban Macrocell (UMa), and Urban Microcell (UMi) models, which encompass a wide range of real-world propagation conditions.

\item Deficiencies in Task Versatility. Current CFMs are predominantly designed for either generation-oriented tasks (e.g., channel reconstruction/prediction) or perception-oriented tasks (e.g., user positioning). There is a lack of a unified framework that can simultaneously predict CSI and utilize it for environmental sensing, thereby restricting the holistic optimization of wireless systems.

\item High Resource Requirements and Limited Transferability. Training and finetuning these models demand substantial computational resources. Moreover, the absence of zero-shot learning capabilities hinders their ability to adapt to new scenarios without additional data.

\end{enumerate}

To address the aforementioned limitations , we propose a novel CFM, CSI-MAE. Our approach begins with simulating a large-scale dataset that spans multiple configurations and incorporates various propagation models for pretraining. By adopting a masked channel modeling strategy during pretraining, CSI-MAE learns robust representations of wireless channels. Additionally, by freezing the pretrained model and finetuning only a lightweight task decoder, CSI-MAE achieves competitive performance with significantly reduced computational resources. Notably, the model showcases remarkable zero-shot performance across different frequencies.

This paper's contributions can be summarized as follows:

\begin{enumerate}
\item \textbf{Generalized Channel Foundation Model}: We propose CSI-MAE, a CFM that is trained on datasets generated by standardized 3GPP statistic channel models. This approach transcends the reliance on scene-specific data, enabling  reusable and cross-scenario generalization.
\item \textbf{Universal Representation for Integrated Sensing and Communication}: CSI-MAE enables concurrent CSI perception and generation, extracting highly generalizable features. Rigorous validation across channel extrapolation, feedback, and user positioning tasks demonstrates its effectiveness in wireless-channel processing.
\item \textbf{Parameter-Efficient and Transferable Adaptation}: Through the finetuning only a lightweight decoder, CSI-MAE achieves comparable or superior performance to supervised models while significantly reducing training costs. Under full-parameter finetuning, our model outperforms previous methods, reaching the state-of-the-art and achieving significant improvements in accuracy. Moreover, CSI-MAE demonstrates remarkable zero-shot transferability, showcasing capabilities comparable to supervised methods in cross-scenarios.
\end{enumerate}

\section{Proposed Framework}

\subsection{Model Architecture}
\begin{figure*}[htbp]
  \centering
  \includegraphics[width=0.8\linewidth]{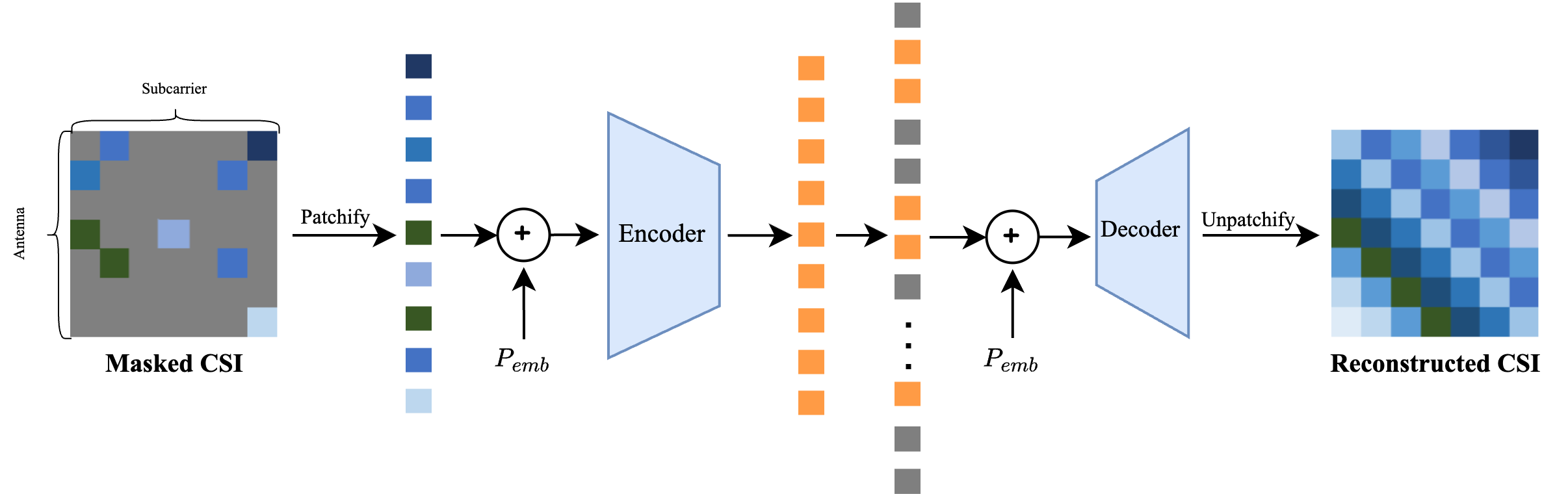}
  \caption{Our proposed CSI-MAE architecture. }
  \label{fig:architecture}
\end{figure*}

CSI-MAE follows the MAE\cite{mae} framework, adapted to the structural characteristics of CSI data. As illustrated in Fig.~\ref{fig:architecture}, it adopts an asymmetric encoder--decoder architecture, where the CSI input is treated as a two-channel image representing its real and imaginary components. A Vision Transformer (ViT)-based \cite{vit} encoder processes only a subset of visible patches, while a lightweight decoder reconstructs the complete channel from the latent features and mask tokens. This asymmetric design ensures high computational efficiency and scalability during self-supervised pretraining.

\textbf{Masking Strategy.} 
The input complex-valued CSI matrix $H$, defined over the antenna and subcarrier domains, is divided into non-overlapping patches \( \mathbf{h}_i \)  and serialized into a 1D token sequence. A high masking ratio (75\%) is applied via uniform random sampling, allowing only 25\% of visible patches to enter the encoder. This design encourages the model to infer the global structure of the channel, rather than rely on local interpolation. Using a low masking ratio may cause the model to find short cuts, enabling it to recover the matrix simply through interpolation without learning general features.

\textbf{Positional Embedding.}
The positional embedding \(P_{\text{emb}}\) preserves the 2D spatial correlation structure of the channel matrix. Specifically, for each spatial index \( (i, j) \) along the \( t \)-th dimension, the embedding components are defined as:

\begin{align}
P_{\text{emb}}^{t}[i,2j] &= 
\sin\!\left(\frac{\text{pos}_i^{t}}{10000^{\frac{2j}{D_{\text{emb}}^{t}}}}\right), \\
P_{\text{emb}}^{t}[i,2j+1] &= 
\cos\!\left(\frac{\text{pos}_i^{t}}{10000^{\frac{2j}{D_{\text{emb}}^{t}}}}\right),
\end{align}
where \( \text{pos}_i^{t} \) is the discrete spatial index and \( D_{\text{emb}}^{t} \) denotes the embedding dimension along axis \( t \). These fixed embeddings are added element-wise to the patch sequence before being processed by the Transformer encoder and decoder layers.

\textbf{CSI-MAE Encoder.} 
The encoder operates on the visible patch sequence using a standard ViT architecture. To retain the spatial relationships across antennas and subcarriers, we incorporate fixed two-dimensional sine--cosine positional embeddings into the patch embeddings,enabling the encoder to capture spatial--frequency dependencies within \(H\). Additionally, a learnable class token \( \text{[CLS]}\) is prepended to the sequence of patch embeddings. This special token serves as a global representation of the entire input sequence, which can be used later for tasks such as classification or as a summary of the encoded information. The encoded latent representation can be formulated as:

\begin{equation}
H_{\text{emb}} = f_{\text{enc}}\left([\text{CLS};H_{\text{vis}}] + P_{\text{emb}}\right),
\end{equation}
where \(H_{\text{vis}}\) denotes the embeddings of visible patches and \(f_{\text{enc}}(\cdot)\) is the encoder, and \( [\text{CLS};H_{\text{vis}}] \) indicates concatenating the learnable class token with the visible patch embeddings.

\textbf{CSI-MAE Decoder.} 
The decoder reconstructs the masked channel patches using both the latent representations and a shared learnable mask token placed at missing positions. Positional embeddings play a crucial role here as well. They help the decoder to understand the relative position of each element in the sequence, which is essential for accurately reconstructing the masked patches. Without positional embeddings, the decoder would lack the spatial context needed to recover the original channel values correctly. Similar to the encoder, class token is also included in the input to the decoder. This allows the decoder to maintain a holistic view of the input sequence and use the global information carried by the class token during the reconstruction process. After reintroducing positional embeddings, the full sequence is processed by a lightweight Transformer decoder followed by a linear projection head to predict the original channel values. The decoder is shallower and narrower than the encoder, focusing on local reconstruction rather than global semantics.

\begin{figure}[tbp]
\centering
\includegraphics[width=\linewidth]{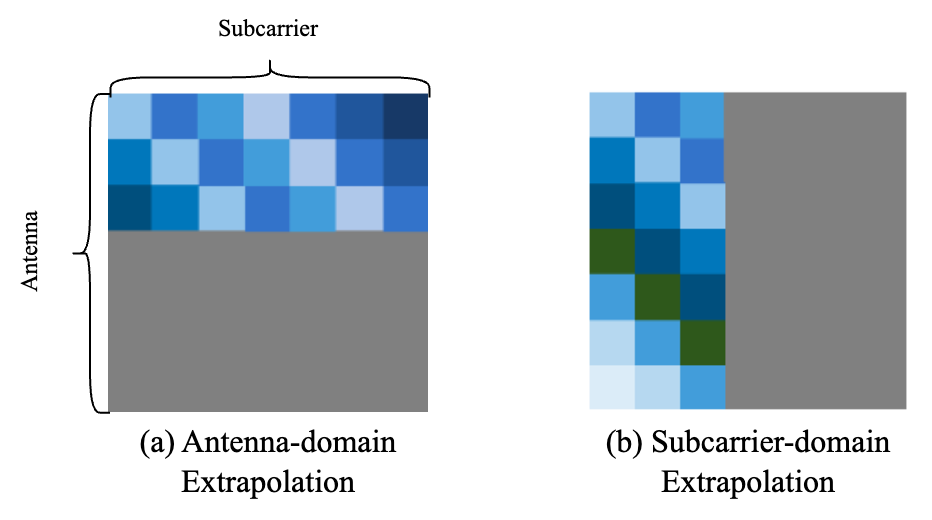}
\caption{Input for antenna-domain and subcarrier-domain extrapolation.}
\label{fig:extrapolation}
\end{figure}

\textbf{Training Objective.} 
The model is optimized by minimizing the Mean Squared Error (MSE) calculated exclusively for masked patches, ensuring that unmasked regions do not contribute to the loss function. The MSE is defined as follows:

\begin{equation}
\mathcal{L}_{\text{MSE}} = \frac{1}{N} \sum_{i=1}^{N} \| \mathbf{h}_i - \mathbf{\hat{h}}_i \|^2,
\end{equation}
where \( N \) denotes the number of masked patches, and \( \mathbf{h}_i \) and \( \mathbf{\hat{h}}_i \) represent the input and reconstructed channel patches, respectively. 
This masked reconstruction objective encourages the model to learn meaningful latent representations that capture the intrinsic spatial–frequency correlations of CSI.

\subsection{Downstream Tasks}

During the finetuning stage for downstream tasks, we leverage transfer learning to exploit the rich feature representations embedded in the pretrained model, thereby boosting task performance. Specifically, the pretrained CSI-MAE encoder has already acquired the ability to extract general-purpose features from CSI data. These features exhibit broad applicability across a variety of CSI-based applications. By appending a lightweight task decoder and subsequently finetuning with labeled data, CSI-MAE can be optimized for specific tasks.

\textbf{Channel Extrapolation and Feedback.} Channel extrapolation aims to predict the unknown CSI by leveraging partial CSI, addressing the issue of incomplete CSI acquisition in practical communication scenarios and enhancing system performance. Specifically, we considers extrapolation in the antenna domain and subcarrier domain. As illustrated in Fig.~\ref{fig:extrapolation}, we apply a 50\% masking ratio to simulate unknown CSI for prediction. Channel feedback, on the other hand, compresses the input CSI and then reconstructs it. 

Both channel extrapolation and feedback are formulated as channel reconstruction tasks, where the goal is to recover the complete channel matrix from partial or compressed observations. 
In extrapolation, only a subset of \(H\) is visible, while in feedback, the full \(H\) is first encoded into a compact latent representation for efficient transmission. 
In both cases, the reconstruction follows a unified forward process:

\begin{equation}
\hat{H} = f_{\text{dec}}\!\left(f_{\text{enc}}\!\left(H_{\text{in}}\right)\right),
\end{equation}
where \(H_{\text{in}}\) denotes the input and \(f_{\text{dec}}(\cdot)\) is the task decoder.  
The learning objective minimizes the reconstruction discrepancy between prediction and ground truth.

\textbf{User Positioning.}  
User Positioning is defined as a regression task that estimates the spatial coordinates of the user equipment. CSI-MAE extracts a latent representation from the input CSI, and a lightweight linear regression head predicts the 2D coordinates based on the class token \( \text{[CLS]}\), which serves as a global representation:

\begin{equation}
\hat{\mathbf{p}} = W_{\text{pos}} \cdot f_{\text{enc}}\!\left(H_{\text{in}}\right)_{\text{[CLS]}} + b_{\text{pos}},
\end{equation}
where \(W_{\text{pos}}\) and \(b_{\text{pos}}\) are learnable parameters.  
The model is optimized with MSE loss.

\section{Experiments}

\subsection{Dataset}
The dataset is generated with Sionna \cite{sionna} under 3GPP TR 38.901\cite{38_901} configurations and covers three representative propagation environments, as shown in Table~\ref{tab:dataset}. Simulations span five carrier frequencies and three subcarrier spacings. In total, the generated dataset contains approximately 1.45 million CSI samples, offering diverse channel conditions across environments, frequencies, and mobility profiles for evaluation. All the data were utilized for pre-training. Five environment-carrier frequency combinations, namely RMa-0.7, RMa-2.4, RMa-3.5, UMa-4.9, and UMi-5, were selected as the test scenarios.

\begin{table}[bp]
\centering
\caption{Simulation Parameter Settings}
\label{tab:dataset}
\resizebox{0.8\columnwidth}{!}{%
\begin{tabular}{cccc}
\hline
\textbf{Parameter}        & \multicolumn{3}{c}{\textbf{Value}}        \\ \hline
Channel Models            & UMi           & UMa         & RMa         \\ \cline{2-4} 
Carrier Frequency (GHz)   & \multicolumn{3}{c}{0.7, 2.4, 3.5, 4.9, 5} \\
Subcarrier spacings (KHz) & \multicolumn{3}{c}{15, 30, 60}            \\
Number of Subcarriers     & \multicolumn{3}{c}{256}                   \\
BS antenna                & \multicolumn{3}{c}{8 $\times$ 8}          \\
BS Height (m)             & 10            & 25          & 35          \\
UE antenna                & \multicolumn{3}{c}{2 $\times$ 2}          \\
UE Height (m)             & \multicolumn{3}{c}{1.5}                   \\
UE Velocity (m/s)         & \multicolumn{3}{c}{{[}0, 27.78{]}}        \\ \hline
\end{tabular}%
}
\end{table}

\begin{figure}[tbp]
  \centering
\includegraphics[width=0.8\linewidth]{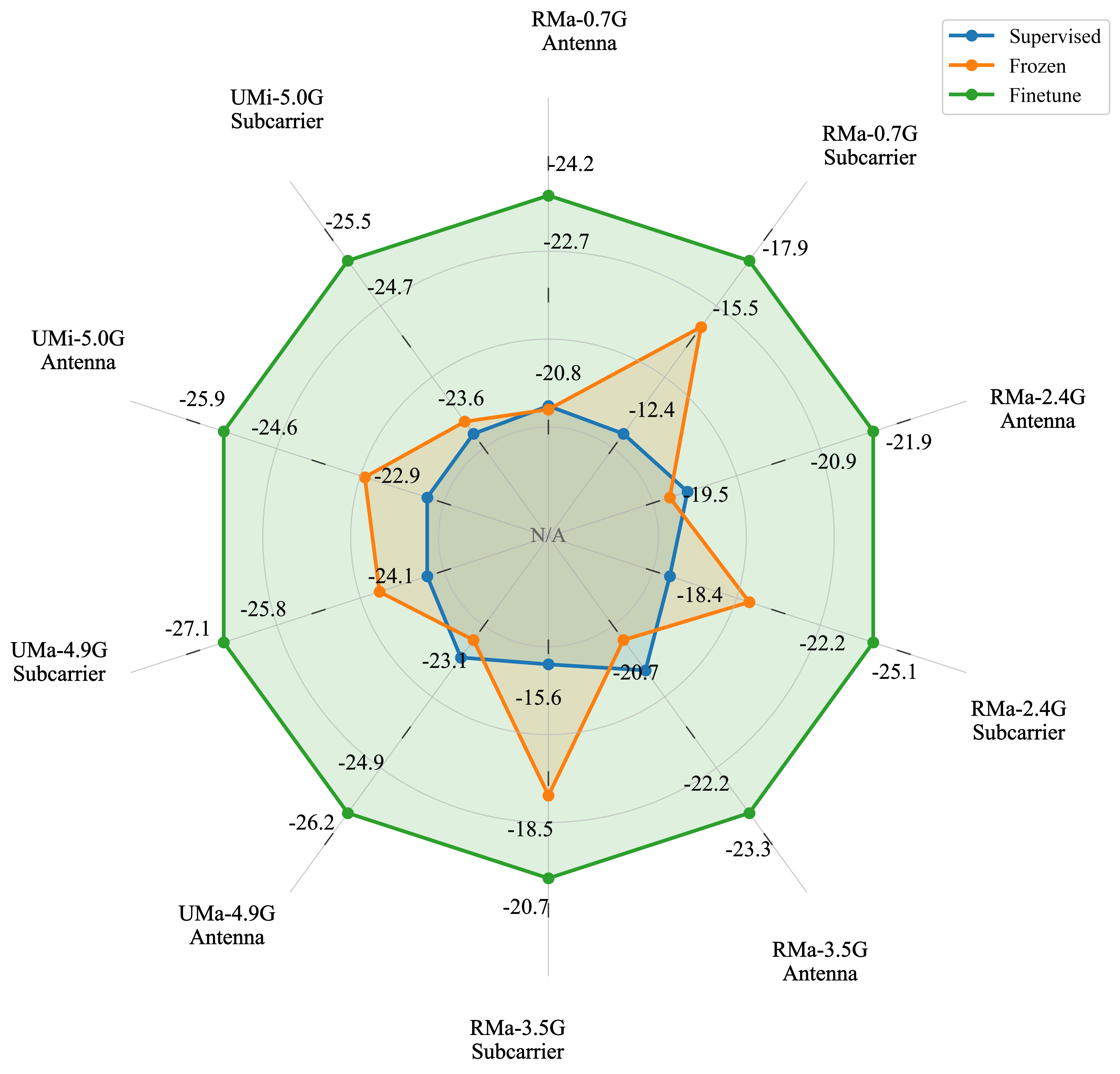}
  \caption{The chart compares performance across scenarios  and antenna/subcarrier-domain extrapolations.}
  \label{fig:extrapolation_radar}
\end{figure}

\subsection{Results analysis}
For pretraining, ViT-Base\cite{vit} architecture, adapted for two channel input, was employed as the CSI-MAE encoder.
Normalized Mean Squared Error (NMSE) are used as the primary evaluation metric. To provide a comprehensive com parison framework, a supervised baseline model (\textit{Supervised}) is incorporated into the experimental setup. This baseline model shares the identical network architecture backbone but is trained without leveraging any pretrain. We explore two transfer learning approaches. In the \textit{Frozen} method, we freeze the CSI-MAE encoder and train only a lightweight task decoder. The \textit{Finetune} approach, conversely, adapts the whole model end-to-end for the target dataset.

\textbf{Channel Extrapolation.} Fig.~\ref{fig:extrapolation_radar} reveals that our CSI-MAE significantly outperforms the supervised baseline in the channel extrapolation task across all five scenarios. Although both approaches share an identical encoder architecture, their core difference lies in the training paradigm. The supervised model, trained solely on local data from each scenario, develops specialized but limited representations. In contrast, the CSI-MAE encoder, pretrained on a broad corpus of data, acquires universal CSI representation that grants it superior generalization capabilities across diverse scenarios.

This is further evidenced by the \textit{Frozen} experiment. Even with a frozen encoder and only training a lightweight transformer decoder, the model achieves performance on par with or even exceeding the supervised baseline. This result strongly validates the power of the learned general-purpose features and highlights the practical value of our CSI-MAE, particularly for applications in data-constrained settings.

\textbf{Channel Feedback.}
As shown in Table \ref{tab:feedback}, our CSI-MAE also demonstrates exceptional performance on the channel feedback task. Similar to channel extrapolation, this task is fundamentally about channel reconstruction. Consequently, the universal CSI features learned during the pretrain stage are highly effective. This is evidenced by the \textit{Frozen}, which, despite having a frozen encoder, consistently outperforms the supervised baseline across all scenarios. By finetuning the entire model, \textit{Finetune} further improves performance, achieving the best results. This again validates the power of CSI-MAE in providing a robust starting point for downstream tasks, proving its strong generalization capabilities for different tasks.

\begin{table}[]
\centering
\caption{Performance comparison of CSI-MAE for channel feedback.}
\begin{tabular}{cccccc}
\hline
\multirow{2}{*}{\textbf{Method}} & \multicolumn{5}{c}{\textbf{NMSE (dB)}}                                                      \\ \cline{2-6}
                           & RMa-0.7 & RMa-2.4 & RMa-3.5 & UMa-4.9 & UMi-5  \\ \hline
Supervised                    & -35.14           & -43.55           & -38.73           & -44.08           & -48.86          \\
Frozen                    & -36.18           & -44.07           & -40.83           & -44.88           & -49.01          \\
Finetune                    & \textbf{-37.11}  & \textbf{-48.25}  & \textbf{-43.31}  & \textbf{-46.52}  & \textbf{-50.76} \\ \hline

\label{tab:feedback}

\end{tabular}

\end{table}

\begin{figure}[tbp]
  \centering
  \includegraphics[width=\linewidth]{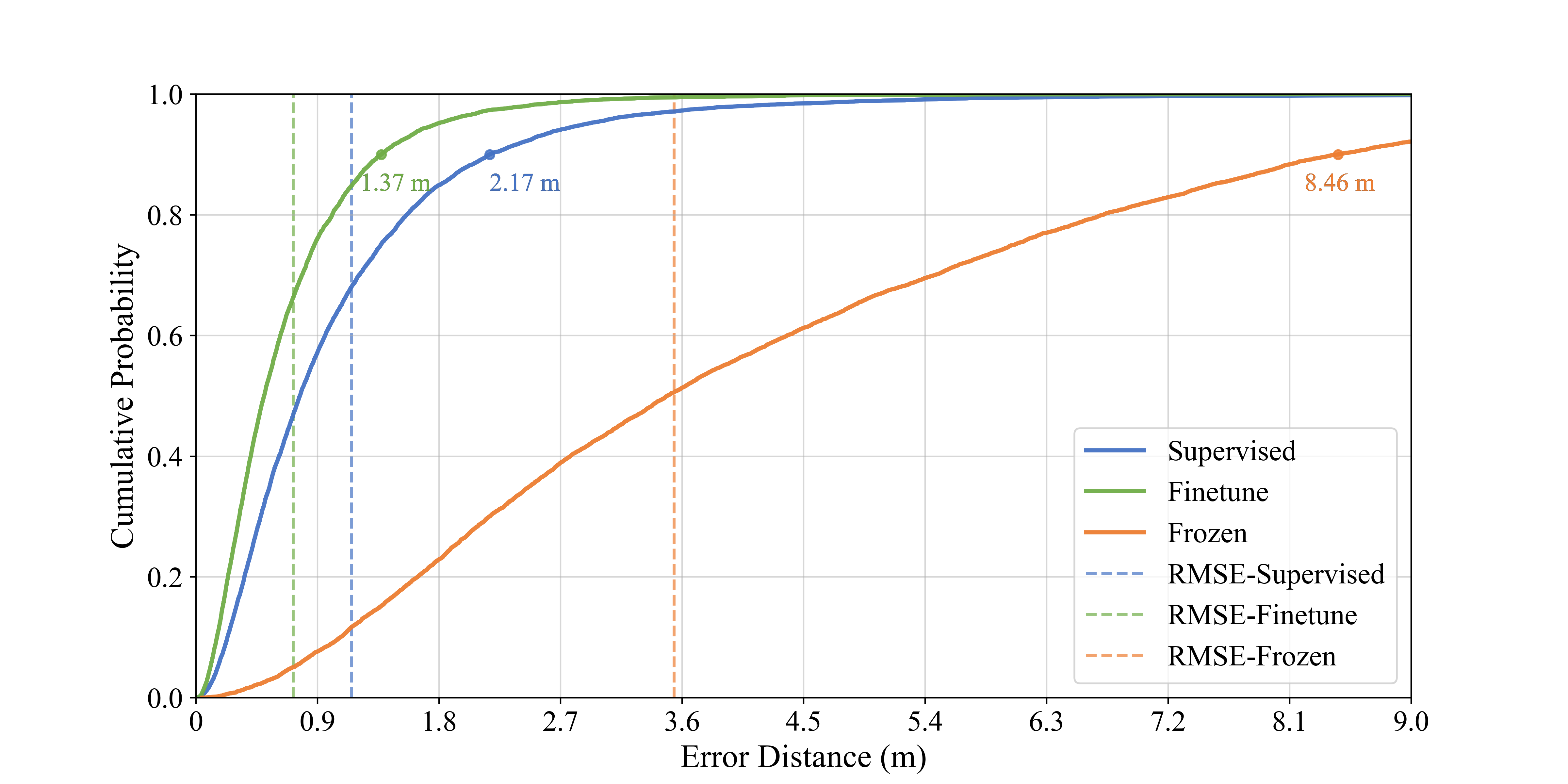}
  \caption{Performance comparison of CSI-MAE for the positioning in CDF.}
  \label{fig:cdf}
\end{figure}

\textbf{User Positioning.} The positioning results, as shown in Fig.~\ref{fig:cdf}, reveal key insights about our CSI-MAE. The \textit{Finetune} configuration excels, with 90\% of errors within 1.37 m and an RMSE of 0.718 m, outperforming the supervised baseline (2.17 m, 1.151 m). Conversely, the \textit{Frozen} configuration underperforms, with 90\% of errors below 8.46 m and an RMSE of 3.539 m. This disparity stems from two factors: First, the decoder used for the positioning task is only a single linear layer, offering limited learning capacity. More importantly, positioning is a different regression problem compared to masked channel modeling. It requires the model to map high-dimensional CSI data to low-dimensional spatial coordinates, demanding higher-level semantic information that differs from those learned for structural reconstruction during pretraining.

Further, we compared MDA\cite{mda}, LWM\cite{lwm}, CSI-CLIP\cite{csi-clip}, and CSI-MAE on the deepmimo-based dataset\cite{deepmimo} introduced by LWM in Table~\ref{tab:pos}. Given the differences in the configurations of the transmit and receive antennas, subcarriers, etc., across datasets, we employed bilinear interpolation to align the input size of pretraining. The results show that CSI-MAE outperforms the existing LWM even when only \textit{Frozen}. It even surpasses the MDA, which is specifically designed for user positioning. Performance is further enhanced by \textit{Finetune}, reducing mean error more than 10 times. However, CSI-CLIP which is pretrained on the deepmimo achieves the best performance in three scenarios.

\begin{table}[]
\centering
\caption{Performance comparison of state-of-the-art on user positioning task with mean distance errors(m).}
\label{tab:pos}
\resizebox{\columnwidth}{!}{%
\begin{tabular}{ccccc}
\hline
\multirow{2}{*}{\textbf{Method}} & \multicolumn{4}{c}{\textbf{Scenarios}}                        \\ \cline{2-5} 
                                 & Chicago       & San Diego     & Charlotte     & Outdoor       \\ \hline
MDA\cite{mda}                             & 48.75         & 285.90        & 295.45        & 3.75          \\
LWM\cite{lwm}                             & 69.45         & 381.58        & 788.11        & 16.84         \\
CSI-CLIP\cite{csi-clip}                         &  \textbf{3.95}         & \textbf{6.79}        & 7.75        & \textbf{0.45}          \\
CSI-MAE (Frozen)                 & 14.29         & 15.35         & 20.69         & 0.93          \\
CSI-MAE (Finetune)               & 5.11 & 9.59 & \textbf{7.18} & \textbf{0.45} \\ \hline
\end{tabular}%
}
\end{table}

\subsection{Zero-shot Performance}

\begin{figure}
    \centering
    \includegraphics[width=\linewidth]{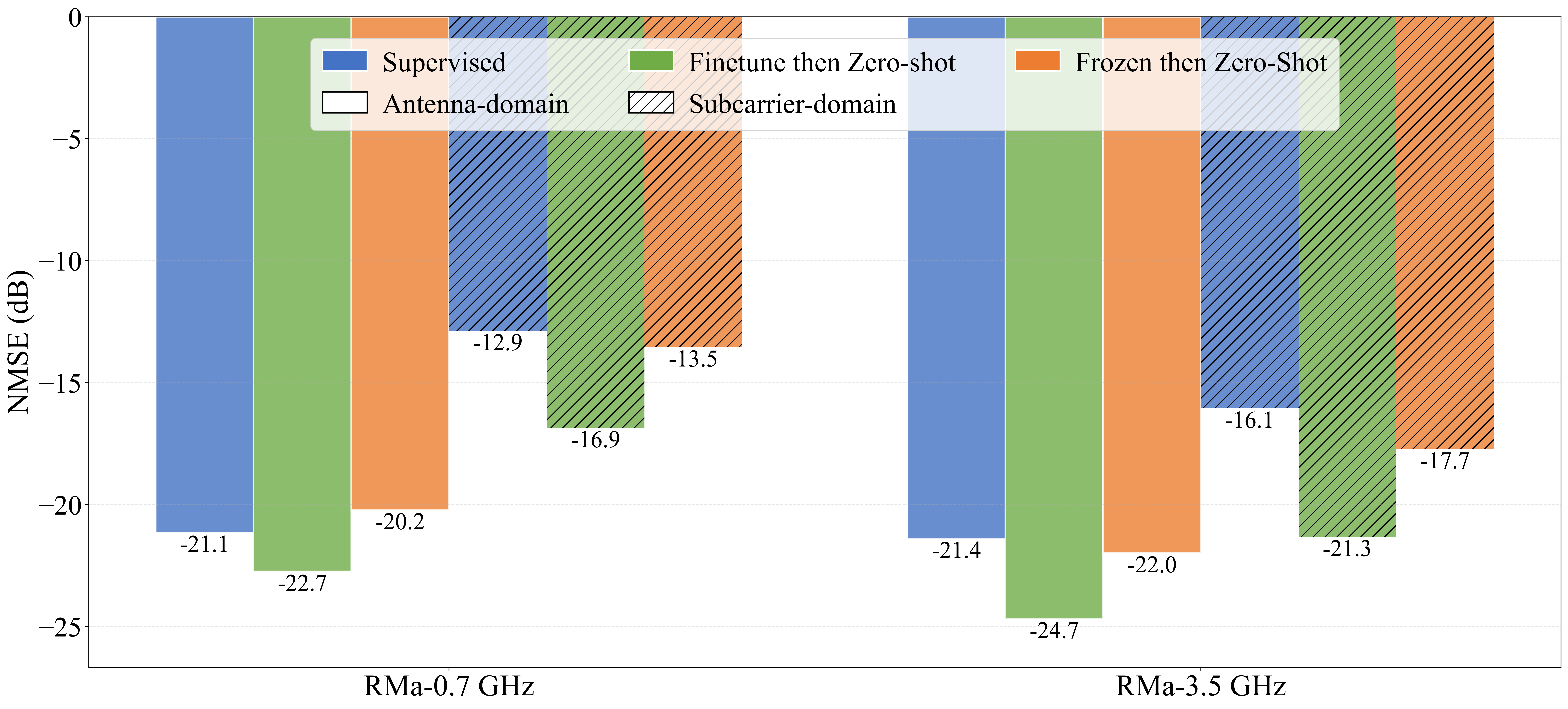}
    \caption{Zero-shot performance comparison on channel extrapolation. The model is fine-tuned on RMa-2.4 and tested on RMa-0.7 and RMa-3.5.}
    \label{fig:zeroshot_extrapolation}
\end{figure}

To further validate the generalization capability of our CSI-MAE, we conducted a zero-shot experiment. A model fine-tuned exclusively on the RMa-2.4 scenario was directly applied to the RMa-0.7 and RMa-3.5 scenarios without any additional training. As illustrated in Fig.~\ref{fig:zeroshot_extrapolation}, our CSI-MAE significantly outperforms the supervised baseline, which was trained specifically on the target data. This remarkable zero-shot performance underscores the model's ability to learn universal, transferable CSI features, enabling it to adapt to entirely new environments and frequency bands effectively.

\subsection{Scaling Law}

Scaling law is essential for improving model performance and generalization. In our research, grasping its influence on the CSI-MAE framework is key to optimization. To fully verify CSI-MAE's scalability, we conducted experiments on data and model scaling.

\textbf{Data Scaling.} We pretrain the model on a partial dataset containing 0.48~M samples restricted to a single subcarrier spacing with the channel extrapolation tasks. As shown in Table~\ref{dataset_scaling}, CSI-MAE which is pretrained on more diverse data achieves superior performance, especially in the \textit{Frozen} on the antenna-domain extrapolation, with a gain of up to 62.7\%. This indicates that large-scale  pretrain data enables the model to learn more generalizable channel representations, demonstrating the framework’s strong data scalability.

\textbf{Model Scaling.} We compared the performance of ViT-Base and ViT-Large for the CSI-MAE's encoder, both pretrained on the full dataset, on the channel extrapolation task. The ViT-Large, which has a deeper and wider architecture than ViT-Base, was evaluated to examine the effect of model capacity. As shown in Table~\ref{model_scaling}, the effectiveness of model scaling is highly dependent on the data dimension. In the subcarrier dimension, the ViT-Large demonstrates a significant advantage, with its finetuned performance improving by 54.8\%, indicating that larger models can more effectively capture the complexities of the frequency domain. However, in the more structured antenna domain, the performance of ViT-Large degrades, with a drop as high as 109.1\%. This suggests that for the antenna dimension, which has a stronger spatial structure, a larger model capacity does not translate to a performance advantage and may even lead to degradation due to over-parameterization. This indicates that the current model training strategy has limitations and that specialized designs accounting for the physical antenna array geometry are needed.

\begin{table}[tbp]
\centering
\caption{Performance comparison of CSI-MAE pretrained on a partial dataset and a full dataset for Channel Extrapolation}
\begin{tabular}{ccccc}
\hline
\multirow{2}{*}{\textbf{Domain}} & \multirow{2}{*}{\textbf{Method}} & \multicolumn{3}{c}{\textbf{NMSE (dB)}} \\ \cline{3-5} 
                              &                                  & Partial-0.48M  & Full-1.45M & $\Delta$ \\ \hline
\multirow{2}{*}{Antenna}      & Frozen                       & -17.43         & -21.83     & \textcolor{red}{+62.7\%}  \\
                              & Finetune                          & -23.69         & -24.31     &  \textcolor{red}{+12.2\%}  \\ \hline
\multirow{2}{*}{Subcarrier}   & Frozen                       & -18.85         & -21.04     &  \textcolor{red}{+31.2\%}  \\
                              & Finetune                          & -21.41         & -23.28     &  \textcolor{red}{+27.3\%}  \\ \hline
\multicolumn{5}{l}{$^{\mathrm{a}}$The improvement $\Delta$ is calculated on the linear NMSE values.}
\label{dataset_scaling}
\end{tabular}
\end{table}

\begin{table}[tbp]
\centering
\caption{Performance comparison of ViT-Base and ViT-Large for CSI-MAE's encoder for Channel Extrapolation}
\begin{tabular}{ccccc}
\hline
\multirow{2}{*}{\textbf{Domain}} & \multirow{2}{*}{\textbf{Method}} & \multicolumn{3}{c}{\textbf{NMSE (dB)}} \\ \cline{3-5} 
                              &                                  & ViT-Base    & ViT-Large   & $\Delta$   \\ \hline
\multirow{2}{*}{Antenna}      & Frozen                       & -21.83      & -18.60      & \textcolor{blue}{-109.1\%}   \\
                              & Finetune                          & -24.31      & -23.04      & \textcolor{blue}{-34.0\%}    \\ \hline
\multirow{2}{*}{Subcarrier}   & Frozen                       & -21.04      & -21.52      & \textcolor{red}{+10.1\%}    \\
                              & Finetune                          & -23.28      & -26.75      & \textcolor{red}{+54.8\%}    \\ \hline
\multicolumn{5}{l}{$^{\mathrm{a}}$The improvement $\Delta$ is calculated on the linear NMSE values.}    
\label{model_scaling}
\end{tabular}
\end{table}

\section{Conclustion}

In conclusion, we proposed CSI-MAE, a Channel Foundation Model trained  on 3GPP-defined statistical channel models using  masked autoencoder. CSI-MAE breaks free from the constraints of scene-specific datasets, achieving standardized, reusable, and cross-scenario generalization, which can extract highly generalizable features, demonstrating effectiveness across multiple wireless-channel processing tasks. Additionally, by finetuning only a lightweight task decoder, CSI-MAE significantly reduces training costs while maintaining competitive or superior performance compared to fully supervised models and exhibits exceptionally robust zero-shot transferability. Notably, we have rigorously validated the scaling law from both data and model dimensions. As the data scale expands and the model parameter count increases, CSI-MAE demonstrates progressively more potent performance. Our work represents a significant step forward in advancing wireless communication technology, offering a more efficient, versatile, and adaptable approach to leveraging Channel Foundation Models.

\bibliographystyle{IEEEtran}
\bibliography{bibfile}

\end{document}